\documentclass{aa}
\usepackage{graphicx}
\usepackage{rotating}
\usepackage{txfonts}
\usepackage{natbib}
\bibpunct{(}{)}{;}{a}{}{,}
\newcommand{\grb}{\object{GRB 011121} }
\newcommand{\xrf}{\object{XRF 011030} }

\begin{document}

\authorrunning{A. Galli \& L. Piro}
\titlerunning{Flaring in XRF 011030}

%\thesaurus{03(11.01.2; 11.02.1; 11.02.2 PKS~B1830$-$211; 11.17.3)}
\title{Long-term flaring activity of XRF 011030 observed with BeppoSAX}
\author{A.~Galli\inst{}
\and L.~Piro\inst{} } \institute{Istituto Astrofisica Spaziale e
Fisica Cosmica - Sezione di Roma, INAF, via del Fosso del
Cavaliere, 100-00113 Roma, Italy}
\offprints{alessandra.galli@rm.iasf.cnr.it}
\date{Received ....; accepted ....}
\markboth{A. Galli \& L. Piro: The temporal and spectral analysis of XRF 011030 }
{A. Galli \& L. Piro:  XRF 011030 }
\abstract{ We present the spectral and temporal analysis of the
X-ray flash \xrf observed with BeppoSAX. This event is characterized
by a very long X--ray bursting activity that lasts about 1500 s,
one of the longest ever observed by BeppoSAX. In particular, a 
precursor and a late flare are present in the light curve.

We connect the late X--ray flare observed at about 1300 s to the
afterglow emission observed by Chandra and associate it with the
onset of the afterglow emission in the framework of external shock
by a long duration engine activity.
We find that the late X-ray flare and the broadband afterglow data,
including optical and radio measurements, are consistent either with
a fireball expanding in a wind environment or with a jetted fireball
in an ISM.

}
\maketitle

\section{Introduction}

Prompt and afterglow emission in GRB show different temporal and
spectral properties and are usually attributed to different mechanisms,
more specifically, internal and external shocks, respectively.
During the prompt emission, the spectrum is hard and shows
strong spectral evolution from hard to soft. However, the
afterglow emission is much softer, and its spectral index remains
roughly constant during the whole emission from early to
late observations \citep{frontera00,piro02}. The
transition from one regime to the other takes place from a few
tens to a few thousand seconds after the burst. In this phase, a
variety of temporal behaviour is observed in different bursts, 
likely due to the contribution of both prompt and afterglow emission.  
Most intriguing is the presence of X-ray flares, such as that observed 
in \xrf.

Indeed, several bursts observed by BeppoSAX showed the
presence of X--ray flares from tens of seconds (e.g., \object{GRB
970228} \citep{frontera} and \object{GRB 980613} \citep{soffitta}) 
to several minutes (e.g., \grb and X-Ray Rich \object{XRR 011211} 
\citep{piro05}) after the burst. These flares have a soft spectrum
consistent with that of the late afterglow. Furthermore, they
connect with the late afterglow emission with a power law $F
\propto (t-t_0)^{-\alpha}$. For early ($<$ 1 minute) X--ray flares,
the origin of the time $t_0$ is consistent with the onset of the
prompt emission. For later ($\gtrsim$ 100 s) X--ray flares, $t_0$
needs to be shifted to the onset of the flare \citep{piro05}.

More recently, X--ray flares taking place in a similar time
period have been observed by Swift \citep{swift} in a larger
number of bursts, about one half of the sample \citep{obrien}
\footnote{We notice that flares and/or re-brightenings are 
also taking place on longer time scales. In the present paper, 
we focus on flares from times of minutes up to $\sim$ 1000 s,                      
i.e., on a time scale similar to that observed in \object{XRF 011030}.}.
Swift observations are providing a big advance in the understanding 
of this phenomenon and its possible relationship with the central engine.
In particular, the discovery in \object{GRB 050502B} \citep{burrows}
of a giant flare, $\sim$ 700 s after the trigger, with an energy 
comparable to that of the burst itself, suggests that the central
engine is undergoing long periods of strong activity.

Swift observations confirm that X-ray flares have a spectrum 
that is globally softer than the prompt emission, i.e., the
peak energy $E_{peak}$ is of the order of a few keV \citep{falcone05}. 
In some cases (\object{GRB 050126}, \object{GRB 050219a}, and 
\object{GRB 050904}), no significant evidence of spectral evolution 
is detected, and the spectrum of the flare is consistent with that
observed in the late afterglow \citep{tagliaferri05,goad05,burrows_rew}. 
The light curve connecting the flare with the late afterglow can be 
reasonably well-fitted by the $(t-t_0)^{-\alpha}$ power law
\citep{tagliaferri05}. However, in other cases (\object{XRF
050406} \citep{romano06}, \object{GRB 050502B}
\citep{burrows,falcone05}, \object{GRB 050421}, \object{GRB
050607}, \object{GRB 050730}, and \object{GRB 050724}
\citep{burrows_rew}), hard-to-soft spectral evolution
was observed during the flare.

Several scenarios were proposed to explain the X-ray flare
phenomenon \citep{zhang05}.
\citet{burrows} propose that the central engine releases energy 
for a long time, and internal shocks then produce a long duration
prompt emission. In the framework of the forward-reverse shock
scenario, \citet{fan} have shown that, adopting different values 
for the forward and the reverse shock parameters, the reverse
shock synchrotron radiation can dominate in the X--ray band
producing a flare. In another scenario, late X--ray flares
mark the beginning of the afterglow emission, and they are produced
by a thick shell fireball (long duration engine activity) through
an external shock \citep{piro05}. Very recently, \citet{wu05} have 
shown that X--ray flares can be produced in the context of both 
late internal and late external shocks. They assume that the central 
engine releases energy in two episodes (i.e., an early and a late shell are ejected).
They applied their model to four Swift GRB and found that 
\object{XRF 050406} and \object{GRB 050607} flares
can be explained both with late internal or external shocks.

In this paper, we present a complete analysis of \object{XRF 011030}
observed with BeppoSAX. It shows outburst activities, with
the last detected flare occurring about 1300 s after the burst.
We investigate the origin of this late X--ray flare. As mentioned
above, several models could explain this phenomenon. Here we have
carried out a detailed analysis of the model in which the flare
is produced by the interaction of the fireball with the external
medium. We check if the model can consistently account for
the broadband data -- from radio to X-rays. We then derive the
main parameters of the fireball, including the density profile of
the surrounding medium. We have also tested this model for the   
X--ray flare occurring in \object{GRB 011121}.

We describe the observations of \object{XRF 011030} in Sect.
\ref{data} and perform its temporal and spectral analysis in
Sect. \ref{reduction} and \ref{spectra}, respectively. In
Sect. \ref{interpretation} we discuss late flares in the context
of different variants of the external shock model. In Sect.
\ref{modelapplication} we apply a long duration engine activity (thick
shell) model to the late flare appearing in \xrf and in \grb, and we
explain it as the onset of the afterglow emission. In particular, in
Sect. \ref{break}, we study the \xrf late afterglow emission taking
into account the presence of a break occurring between $10^4$ and
$10^6$ s after the burst.
In Sect. \ref{optical} we use information on \xrf from the
optical and the radio band to further constrain the model. Our
results and conclusions are summarized in Sect. \ref{conclusions}.

\section{Observations and data reduction}
\label{data}

The X-ray flash \object{\xrf} was detected by the BeppoSAX Wide
Field Camera (WFC) no. 1 on October 30th, 2001, at 06:28:02, without
any counterpart in the Gamma Ray-Burst Monitor (GRBM) \citep{gand}.

The peak flux is $7.5\times10^{-9}$erg s$^{-1}$cm$^{-2}$, and the total
fluence of the source between 2 and 26 keV is equal to $1.2\times10^{-6}$
erg cm$^{-2}$, consistent with the typical value observed in the same
range for normal GRB \citep{amati}.

The X-ray afterglow of \xrf was identified by Chandra in a 47 ks
exposure beginning on  November 2001, at 9.73 UT and in a second
one of 20 ks performed on November 2001, at 29.44 UT \citep{harrison}.
The localization of the X--ray afterglow was consistent with
the position of a radio transient \citep{taylor}. The radio source
was detected on November 2001, at 8.80 UT near the centre of the
WFC error circle at (epoch 2000) R.A.=20:43:32.3, Dec.=+77:17:18.9
with an error of $\pm1 \arcsec$. In this paper, we have used the results
of the analysis of the Chandra data performed by \citet{vale}. The
spectrum between 2 and 10 keV is fitted by a power law with a photon
index $\Gamma=1.72^{+0.19}_{-0.20}$ (Table \ref{analysis}).
Several optical observations were carried out, but none of them succeeded
in associating an optical counterpart with \object{XRF 011030}. The tighest
upper limits are R$>$21 at 0.3 days after the burst \citep{vijay} and
R$>$23.6 at 2.7 days after the burst \citep{rhoads01}.

The precise Chandra localization allowed the association of the
burst with a faint irregular blue galaxy observed by the Hubble
Space Telescope and the Keck. The photometric observations of
this galaxy suggest a redshift smaller than $z \sim$3.5, while
the low brightness of the galaxy suggests that $z >$ 0.6 \citep{bloom}.
Since the observations allow us to establish only a wide range of
redshift values, in our analysis we assume $z$=1.\\

\subsection{Temporal analysis of XRF 011030 }
\label{reduction}

\begin{figure*}[!htb]
\centering
\includegraphics[height=14cm,width=10.2cm,angle=-90]{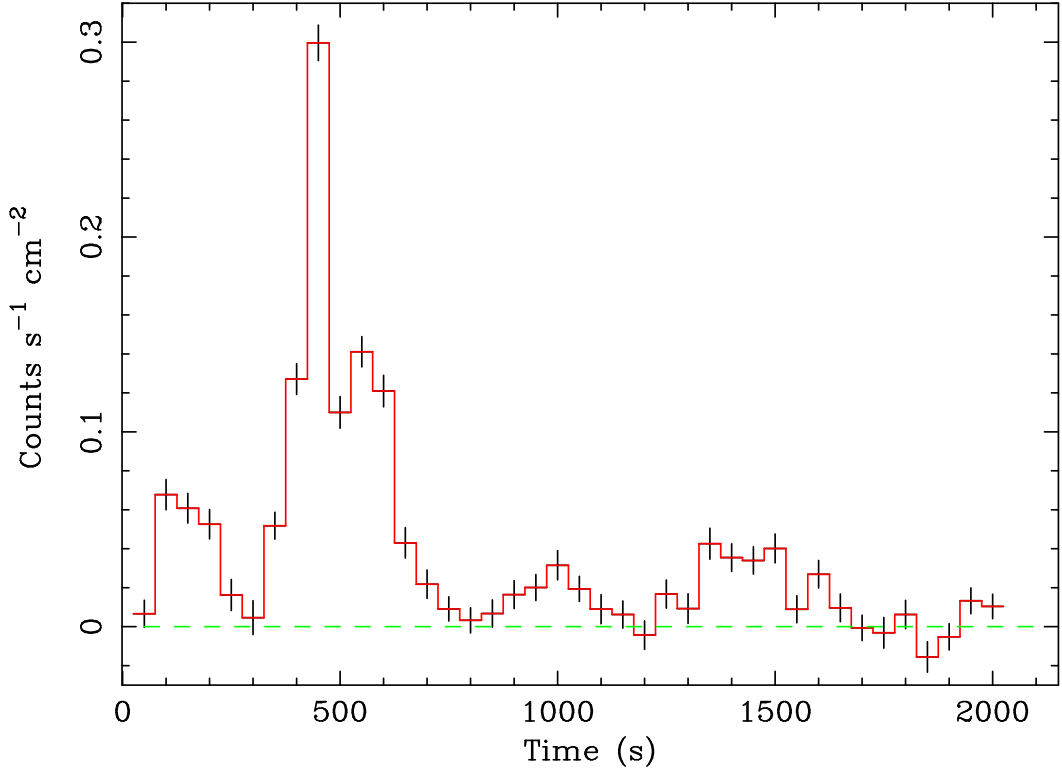}
\caption{ Light curve of \xrf in the BeppoSAX WFC (2-26 KeV) with a temporal
resolution of 50 s. The main pulse (300-700 s) is preceded by a fainter event (0-300 s)
and is followed by a late X--ray flare (1300-1550 s). }
\label{reburst50}
\end{figure*}

We produced background subtracted light curves of \xrf normalized
to the detector effective area exposed to the source.               
The source remained in the field of view (f.o.v.) of the WFC for
about 1 day. A significant source flux above the background level
is detected until $\sim$ 1600 s (see Fig. \ref{reburst50}).
A main pulse, lasting $\sim$ 400 s, starts $\sim$ 300 s after a
fainter preceding event. The main event is also followed by an
X--ray flaring activity, 200 s long, which appears $\sim$ 1300
s after the first pulse. This X--ray flare has duration and flux
similar to the pulse preceding the main event.

As the event is an X--ray flash, its spectrum is soft by definition. 
Thus, we cannot expect to find such substantial spectral differences 
as those find in GRBs between the phases of precursor, prompt emission, 
and late X--ray flare, i.e., with a precursor and late flare markedly 
softer than the prompt emission \citep{piro05}.
In any case, due to the similarities of the bursting activities that preceded and
followed the main pulse in \xrf to those observed in \object{GRB 011121} \citep{piro05},
in the following we refer to these two pulses as precursor and flare, respectively.

\begin{figure}[!htb]
\centering
\includegraphics[width=6.3cm,height=8.75cm,angle=-90]{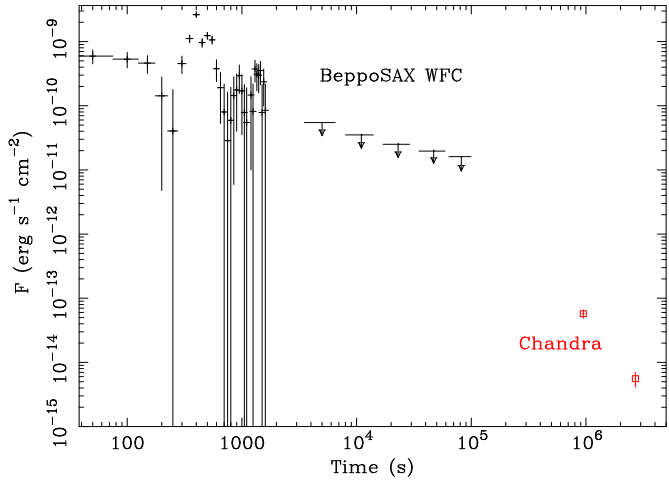}
\caption{\xrf light curve with the BeppoSAX upper limits on the flux fixed at 3$\sigma$
above the background fluctuations. The late afterglow emission detected by Chandra is
shown in red squares. [See the electronic edition for a colour version of this figure.]}
\label{afterglow}
\end{figure}

After about 1600 s, no signal is detected and we can only estimate upper limits on the flux.
The light curve of \xrf with the upper limits between 1600 s and $10^5$ s together with the
late afterglow emission detected by Chandra is shown in Fig. \ref{afterglow}.

\subsection{Spectral analysis}
\label{spectra}

We extracted the spectrum between 2--26 keV from the WFC data. In
the spectral fitting, we tested a simple power law model (with and
without photoelectric absorption), a broken power law model, and a
black-body model. The results of our spectral analysis are
summarized in Table \ref{analysis}. All errors are quoted at 1
$\sigma$ (68\% confidence level).

\begin{figure}[!htb]
\centering
\includegraphics[height=8.8cm,width=6.3cm,angle=-90]{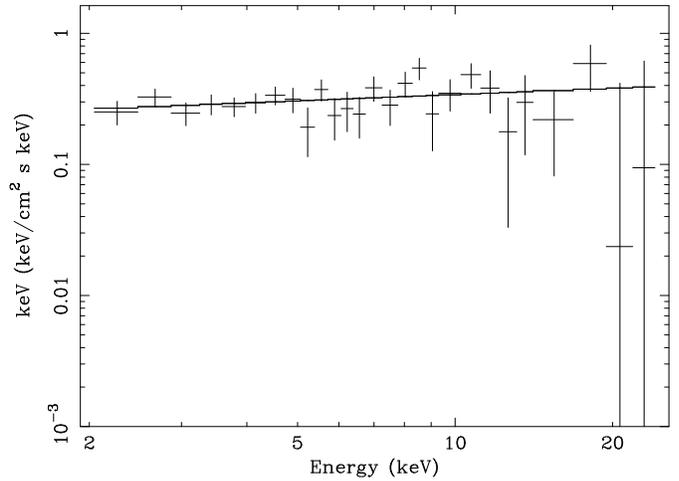}
\caption{ $\nu$F$_{\nu}$ spectrum of the total event.
The solid line represents the fit of \xrf with a simple power law model.}
\label{fit}
\end{figure}

The whole spectrum of \object{XRF 011030}, integrated from 0 s to
1550 s can be described by a simple power law (Fig. \ref{fit}) with a
photon index $\Gamma$=$1.84^{+0.17}_{-0.16}$, consistent with the
simple power law photon index $\Gamma=1.9\pm0.1$ determined by
\citet{heise}. The fit with this model gives $\chi^2_{\nu}=0.83$;
25 degrees of freedom (d.o.f.).
A fit of the spectrum with a broken power law having the photon index
$\Gamma_1$ free to vary and the photon index $\Gamma_2$
fixed to the typical value 2.5 \citep{amati} did not bring a
significant improvement of $\chi^2$. Also, the fit with a power law
with a photoelectric absorption did not bring a significant
improvement of $\chi^2$, and led to an upper limit on the absorption  
column density $N_H<1.5\times10^{23}cm^{-2}$ at z=1. Finally, the fit
with a black-body model gives a $\chi^2_{\nu}$ value greater
than 2, and it can thus be rejected.

We studied the spectral evolution of \xrf by dividing the data into
four intervals: the precursor (from 35 s to 280 s), the first segment 
of the prompt emission (from 280 s to 500 s), the second segment of 
the prompt emission (from 500 s to 1200 s), and the flare (from 1300 s 
to 1550 s). We observed only a marginal spectral variability.

\begin{figure}[!htb]
\centering
\includegraphics[height=8.8cm,width=6.3cm, angle=-90]{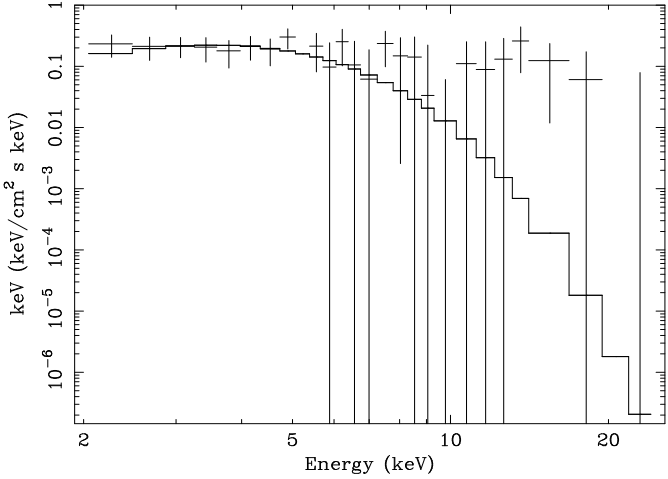}
\caption{$\nu$F$_{\nu}$ spectrum of the precursor. The solid line is
the fit of the precursor with a black-body model.}
\label{bbody}
\end{figure}

The precursor spectrum is well-described by a simple power law with
a photon index $\Gamma$=$2.61^{+0.76}_{-0.61}$, marginally steeper
than the spectrum of the main event. This fit gives $\chi^2_{\nu}$=0.81;
25 d.o.f. The precursor can be also described by a black-body model
with a temperature $kT$=$0.90^{+0.19}_{-0.15}~keV$ (see Fig. \ref{bbody})
and $\chi^2_{\nu}$=0.96; 25 d.o.f. This result is interesting because
there is only one burst, observed by GINGA, whose spectrum is consistent
with a black body \citep{murakami}. However, we cannot discriminate between
these two models.

The first and the second parts of the prompt emission are
both fitted by a simple power law; for the first part, the 
photon index is $\Gamma$=$1.78^{+0.17}_{-0.16}$ and
$\chi^2_{\nu}$=1.24; 25 d.o.f., and for the second one,
$\Gamma$=$1.63^{+0.33}_{-0.30}$ and $\chi^2_{\nu}$=1.48; 25 d.o.f.

Finally,  the late X--ray  flare is also fitted by a simple power
law. Its spectrum is marginally steeper than those of the main
event, as we also found in the case of the precursor:
$\Gamma$=$2.10^{+0.83}_{-0.64}$ with $\chi^2_{\nu}$=1.51; 25
d.o.f.

\begin{table*}[!htb]
\caption{ Results of the spectral analysis of \object{XRF 011030}. The models used are: Power Law (PL),
BroKeN power law (BKN), Power Law plus photoelectric ABSorption (PL+ABS) and Black Body (BB).}
\label{analysis}
\begin{tabular}{c c c c c c c c c c} %9 colonne centrate
\hline % inserisce le righe orizzontali
 name & interval & model & photon & $N_H$ & $E_b$ & kT & $Flux_{2-26~keV}$ & $\chi^2_{\nu}$ & $\nu$ \\  % nomi colonne
    &  (s)  &  & index $\Gamma$ & ($cm^{-2}$, z=1)& (keV) & (keV) & [$erg \cdot cm^{-2}\cdot s^{-1}$] &  &  \\
\hline \hline   % inserisce le singole righe orizzontali
total     & 0-1550 & PL     & $1.84^{+0.17}_{-0.16}$ &       ---           & ---    & ---              & $1.3\cdot10^{-9}$  & 0.83 & 25 \\
event     & --- & BKN    & $1.77^{+0.19}_{-0.23}$ &       ---           & $<11$ & ---                  & $1.3\cdot10^{-9}$  & 0.81 & 24 \\
          & --- & PL+ABS & $1.88^{+0.27}_{-0.14}$ & $<1.5\cdot10^{23}$ &   ---  & ---                  & $1.2\cdot10^{-9}$  & 0.86 & 24 \\
\hline
precursor & 35-280 & PL     & $2.61^{+0.76}_{-0.61}$ &       ---           &   ---  & ---              & $5.8\cdot10^{-10}$ & 0.81 & 25 \\
          & --- & PL+ABS & $2.44^{+2.06}_{-0.44}$ & $<7\cdot10^{22}$    &  ---   & ---                 & $5.5\cdot10^{-10}$ & 0.87 & 24 \\
          & --- & BB     &     ---                &       ---        &  ---   & $0.90^{+0.19}_{-0.15}$ & $3.7\cdot10^{-10}$ & 0.96 & 25 \\
\hline
burst     & 280-500 & PL     & $1.78^{+0.17}_{-0.16}$ &       ---           &  ---   &   ---           & $2.5\cdot10^{-9}$  & 1.24 & 25 \\
part 1    & --- & BKN    & $1.59^{+0.23}_{-0.27}$ &       ---        & $9.1^{+4.0}_{-2.1}$ & ---       & $2.3\cdot10^{-9}$  & 1.08 & 24 \\
          & ---  & PL+ABS & $2.23^{+0.36}_{-0.31}$ & $<4.4\cdot10^{23}$ &   ---   &     ---            & $2.1\cdot10^{-9}$  & 1.16 & 24 \\
\hline
burst     & 500-1200 & PL     & $1.63^{+0.33}_{-0.30}$ &       ---           &  ---   &        ---    & $8.9\cdot10^{-10}$ & 1.48 & 25 \\
part 2    & --- & BKN    & $0.39^{+1.49}_{-0.31}$ &        ---       & $<4.4$ &    ---                 & $7.8\cdot10^{-10}$ & 1.6  & 24 \\
          & ---  & PL+ABS & $1.93^{+0.74}_{-0.53}$ & $<8.6\cdot10^{22}$ &   ---  &      ---           & $8.1\cdot10^{-10}$ & 1.65 & 24 \\
\hline
X-ray late & 1300-1550 & PL     & $2.10^{+0.83}_{-0.64}$ &       ---           &  ---   &    ---       & $4.9\cdot10^{-10}$ & 1.51 & 25 \\
flare     & --- & BKN    & $4.18\pm10$            &       ---           & $<3.0$  &   ---              & $5.8\cdot10^{-10}$ & 1.7  & 24 \\
          & --- & PL+ABS & $2.10^{+3.93}_{-0.64}$ & $<1.3\cdot10^{23}$ &  ---   &  ---                & $4.9\cdot10^{-10}$ & 1.72 & 24 \\
\hline
afterglow$^1$  & $(9.24-9.71)\cdot10^5$ & PL+ABS & $1.72\pm0.20$ & $2.96^{+0.60}_{-0.65}\cdot10^{21}$ & ---  & --- & $5.8\cdot10^{-14}$ & 0.76 & 9 \\
\hline
\end{tabular}
 $^1$From \citet{vale}; the flux is in the 2-10 keV range.
\end{table*}

\section{The late X--ray flare in the context of external shock models}
\label{interpretation}

Among the different models proposed for X-ray flares 
\citep{zhang05}, we choose to analyse in detail some of the models 
based upon an external shock origin, motivated by the spectral similarity 
observed in the flare and afterglow phases, straightforwardly accounted 
for in this scenario. A detailed analysis is carried out to check the 
capability of the model to account for the whole set of broadband data.

In what follows, we first try to explain the late flare of \xrf  as being  due to
external shock in a ''standard'' fireball model (i.e., thin shell case, \citet{sari}),
with a continuous or discontinuous density profile. Since the flare cannot be
described by this model, considering the similarity of \xrf with \object{GRB 011121},
we finally explain it by shifting the origin of time $t_0$ to the instant of the
flare, which corresponds to a thick shell fireball.

\subsection{The Fireball Model: the standard ``thin'' shell case}
\label{standard}

\begin{figure}[!htb]
\centering
\includegraphics[height=8.75cm,width=6.3cm,angle=-90]{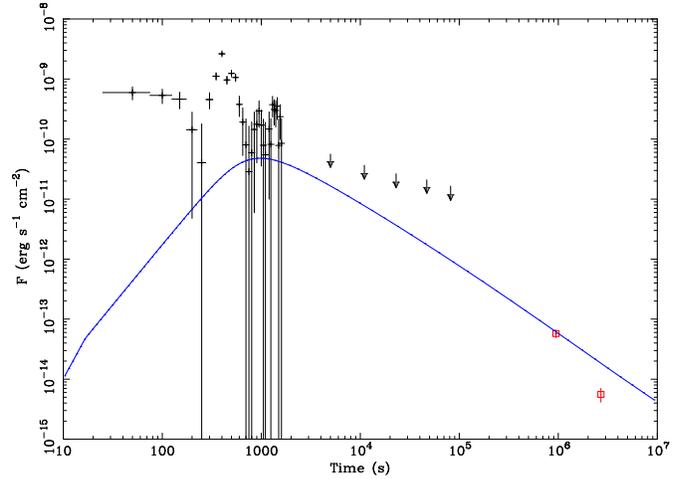}
\caption{X--ray light curve of \xrf for a standard thin
shell fireball expanding in an ISM with $E_{53}=0.03$, $\Gamma_0=45$, 
$n=1$, $\varepsilon_e=0.3$, $\varepsilon_B=0.05$, and $p=2.1$.} 
\label{standardism}
\end{figure}

In a ``standard'' approach, the Fireball Model assumes a thin
shell \citep{sari} that expands with spherical symmetry either in  
a constant density medium or in a wind profile environment. In
this framework, the emitted flux reaches its maximum at the
deceleration radius $r_0$ and then starts to decrease. However, it
does this too slowly to be consistent with the onset and decay of
the flare. This appears clearly in Fig. \ref{standardism}, where
the calculated light curve for a thin shell fireball expanding in
an ISM is shown.

Moreover, we note that a thin shell fireball also fails to explain the
emission at about 1000 s. In such a case, the deceleration time of the fireball
$t_{dec}$ is greater than the duration of the central engine activity $t_{eng}$,
thus prompt and afterglow emission are well separated, and one would expect no
emission between these two phases.

\subsection{Model with a discontinuous density profile}
\label{discontinuity}

A discontinuous density profile can be produced by a variable
activity of wind emission or by interaction of the wind bubble
with the external uniform medium.
When the fireball expands in a medium characterized by a sudden increase
of density, one could expect that a larger number of photons is produced
and the flux increases quickly. This is true only when the electron
cooling frequency $\nu_c$ is greater then the observational frequency
$\nu_{obs}$. On the contrary, when $\nu_{c}<\nu_{obs}$, the
emitted flux is independent of the density profile both during the
slow cooling and the fast cooling regimes \citep{panaitescu00}.
Let us then verify if the cooling frequency could be located above
the X-ray range at the time of the flare.
\begin{flushleft}
A fireball expanding in an ISM decelerates at a time $t_{dec}$ \citep{panaitescu00}:

\begin{equation}\label{eq:tempoism}
t_{dec}\simeq46.7E_{51}^{1/3}n^{-1/3}\Gamma_{0,2}^{-8/3}~s.
\end{equation}

During the deceleration phase (for $t>t_{dec}$), the cooling
frequency $\nu_c$ is given by \citep{panaitescu00}:

\begin{equation}\label{eq:cool}
\nu_c\simeq3.4\times10^{16}E_{51}^{-1/2}n^{-1}\varepsilon_{B,-2}^{-3/2}t_3^{-1/2}~Hz.
\end{equation}
\end{flushleft}

Except for very low values of the Lorentz factor, $\Gamma_0<30$,
the flare occurs during the deceleration phase and the cooling
frequency $\nu_c$ is given by Eq. \ref{eq:cool}. This
equation indicates that, for typical XRF energies and parameter
values, at the time of the flare, $t_3\simeq1$, the cooling
frequency $\nu_c$ could be higher than the observational frequency
$\nu_{obs}$ (that is, in the X--ray band) only for small values
of density, $n<0.07$. During the deceleration phase, $\nu_c$
decreases with time and will pass below the X-ray range at
later times.

\begin{flushleft}
If the fireball starts its expansion in a wind density profile,
the deceleration time $t_{dec}$ is \citep{panaitescu00}:

\begin{equation}\label{eq:tempowind}
t_{dec}\simeq6.67E_{51}A_{*,-2}^{-1}\Gamma_{0,2}^{-4}~s
\end{equation}

and during the deceleration regime

\begin{equation}\label{eq:coolwind}
\nu_c\simeq3.77\times10^{16}E_{51}^{1/2}A_{*,-2}^{-2}\varepsilon_{B,-2}^{-3/2}t_3^{1/2}~Hz.
\end{equation}
\end{flushleft}

Also in this case, for $A_{*}\lesssim 10^{-3}$, $\nu_c$ can be
higher than $\nu_{obs}$ at the flare time. Moreover, now $\nu_c$
increases with time, and therefore the X-ray afterglow emission
will remain sensitive to density variations.
Thus, for a suitable range of parameters, at the
time of the flare the X-ray emission can be sensitive to density.
However, the duration and amplitude of the flare are not consistent
with the kinematic upper limit  recently established by \citet{ioka}
on the flares produced by the interaction of the fireball with density
discontinuities. In particular, if we assume to observe GRB on
axis, this upper limit is \citep{ioka}:

\begin{equation}\label{eq:upperlimit}
\frac{\Delta F_\nu}{F_\nu}\leq \frac{4}{5}f_c^{-1}\frac{F}{\nu
F_\nu}\frac{\Delta t}{t-t_0},
\end{equation}

where  $f_c^{-1}\sim (\nu_i/\nu_c)^{(p-2)/2}$ is the fraction of cooling 
energy and $F/\nu F_{\nu}\sim(\nu_{obs}/\nu_c)^{(p-2)/2}$ for $\nu_c<\nu_{obs}$. 
The flare duration is $\Delta t \simeq 200$ s in \object{XRF 011030},
and the time of the flare occurrence is $(t-t_0)\simeq 1300$ s, where the
time is counted from $t_0$, i.e., in the case of a thin shell, from the
initial trigger. Thus $\Delta t/t \sim$0.15, and with typical parameters
values, Eq. \ref{eq:upperlimit} implies $\Delta F_{\nu}/F_{\nu} \leq$ 0.25,
while from the X--ray data of \xrf we derive $\Delta F_{\nu}/F_{\nu} \sim$ 3.6.

We point out that the case discussed above applies only to
a density discontinuity with a shell geometry. \citet{dermer99}
have shown that a clumpy medium would be able to produce high
variable light curves through external shock if the clouds radius
is very small in comparison to their distance from the central
engine. This process can explain X-ray flares up to thousand of
seconds \citep{dermer05}.

\subsection{Long duration engine activity: the thick shell model }
\label{reburst011030}

In the following, we show how we can describe the flare in the
context of the external shock scenario by shifting the origin of
the time $t_0$ to the onset of the flare.
From a theoretical point of view, the onset of the external shock
depends on the dynamical regime of the fireball that is strictly
related to the ``thickness'' of the shell \citep{sari}, i.e., to
the duration of the engine activity. In fact, a shell is defined
as being thin or thick depending on its thickness $\Delta=ct_{eng}$, where
$t_{eng}$ is the duration of the engine activity, and also on its
initial Lorentz factor $\Gamma_0$. The shell is defined to be
thick if $(E/nm_pc^2)^{1/3}\Gamma_0^{-8/3}<\Delta$ \citep{sari}.
For our purpose, we rewrite the above equation substituting the
deceleration time given in Eq. (3) of \citet{piro05}, obtaining
$t_{eng} \gtrsim t_{dec}$ for the thick shell condition. Most of
the energy is transferred to the surrounding material at $t_{dec}$
for thin shells or at $t_{eng}$ for thick shells. In the latter
case, the peak of the afterglow emission therefore coincides with
$t_{eng}$. Also, the afterglow decay will be  described by a
power-law only if the time is measured starting from the time at
which the inner engine turns off, i.e., $t_0\simeq t_{eng}$.
According to \citet{lazzati}, this should happen when the
central engine releases  most of the energy during the last phase
of its activity.
In this context, the flare would thus be produced by the external
shock caused by an energy injection lasting until the time of the flare
occurrence, i.e., $t_{eng}\approx 1300$s.
The hypothesis of external shock for the flare offers a
straightforward explanation of the spectral similarity with the
late afterglow data.
We also notice that in this model the early afterglow
emission is mixed with the (internal shock) GRB emission \citep{sari}.
In this context, the emission observed at 1000 s can be attributed 
to internal shock while the flare represents the onset of the afterglow emission.

To develop our model, we used the prescriptions of the so-called
standard fireball model \citep{panaitescu00}. They offer analytic
solutions only at distances greater than the deceleration radius
$r_0$. But we are also interested in the so-called coasting and
transition phases \citep{zhang} because we want to study and to
reproduce the shape of the flare, its rise included. As mentioned
above, we have taken into account the thick shell variant by
introducing a time shift $t_0$, i.e., implying that most of the
energy for the external shock is carried at $t_0 \approx t_{eng}$.
We thus numerically solved the basic equations of the fireball
model. The program requires the parameters of the model,
namely the initial value of the Lorentz factor of the relativistic
shell $\Gamma_0$, the energy value in unity of $10^{53}$ erg
$E_{53}$, the electron population index $p$, the fraction of
energy going into relativistic electrons $\varepsilon_e$, the
fraction of energy going in magnetic field $\varepsilon_B$, and the
density of the external medium $n$ (cm$^{-3}$) or $A_*$
(cm$^{-1}$) as input. The density profile is described by the law
$n=3.0\cdot10^{35}A_*r^{-s}$, where in the case of an ISM, $s=0$,
while in the case of a wind profile environment, $s=2$.

When not stated otherwise, we have taken $E_{53}$=0.03,
assuming that all the kinetic energy is converted in $\gamma$-rays 
and that the redshift is $z=1$. From our spectral analysis, we find 
$p=2.1$. To determine the other parameters values, 
we performed a study devoted to understanding how they influence the 
calculated light curve.

We investigated the effect of model parameters on the X-ray light
curve produced by a thick shell fireball with spherical symmetry
expanding in an ISM. The origin of the time, $t_0$, is shifted to 
the instant of the flare, 1300 s.
We show the X--ray flux  between 2-10 keV obtained by numerical
integration of the specific energy flux.

\begin{figure}[!htb]
\centering
\includegraphics[height=8.75cm,width=6.3cm,angle=-90]{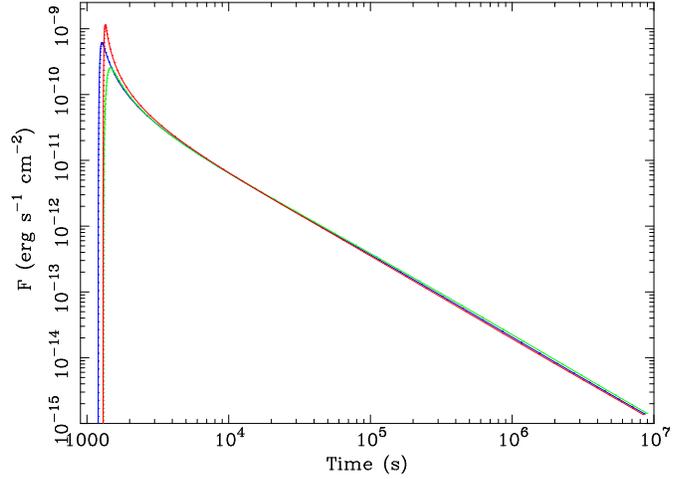}
\caption{Effects of density $n$ of the external medium on the X--ray light
curve with the origin of the time shifted to $t_0$=1300 s.
For the red curve $n$=5 $cm^{-3}$, for the blue curve $n$=1 $cm^{-3}$, and for the green
curve $n$=0.1 $cm^{-3}$. The X--ray light curve peak shows a small increase with $n$.
The other model parameters are $E_{53}$=0.03, $\Gamma_0$=85, $\varepsilon_e$=0.01,
$\varepsilon_B$=0.05, and $p$=2.2.}
\label{density}
\end{figure}

First we find, according to the fact that the X-ray emission is
typically above the cooling frequency, particularly at late times,
that the density $n$ and $\varepsilon_B$ only have a marginal effect
on the normalization of the X-ray light curve.
Figure \ref{density} shows, for example, the effects of the 
density $n$ of the external medium in which the fireball expands.
Differences are appreciable only at early times, when the
observational frequency $\nu_{obs}$ is smaller than the cooling
frequency $\nu_c$.

\begin{figure}[!htb]
\centering
\includegraphics[height=8.75cm,width=6.3cm,angle=-90]{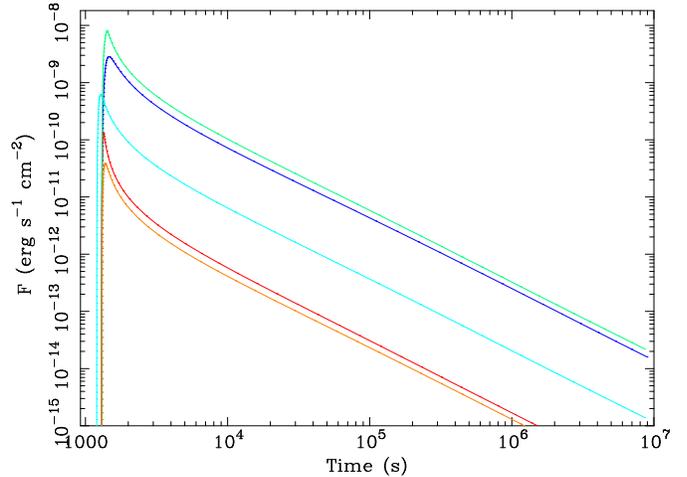}
\caption{ Effects of the energy $E_{53}$ and $\varepsilon_e$ on the
X--ray light curve with the origin of the time shifted to $t_0$=1300
s. For the green curve $E_{53}=0.03$ and  $\varepsilon_e=0.1$, for
the blue curve $E_{53}=0.3$ and $\varepsilon_e=0.01$, for the
light blue curve $E_{53}=0.03$ and $\varepsilon_e=0.01$, for the
red curve $E_{53}=0.003$ and $\varepsilon_e=0.01$, and finally for
the orange curve $E_{53}=0.03$ and $\varepsilon_e=0.001$.
The other model parameters are $n$=1, $\Gamma_0$=85, $\varepsilon_B$=0.05,
and $p$=2.2. Note, as the emitted flux increases with $E_{53}$ and $\varepsilon_e$,
that these parameters have about the same amount of influence in determining
the normalization factor of the X--ray light curve.} \label{energy}
\end{figure}

The effects of the parameters $E_{53}$ and $\varepsilon_e$ are
presented in Fig. \ref{energy}. The normalization of the X-ray
light curve depends mostly on the product of $E_{53}$ and
$\varepsilon_e$, and follows a roughly linear dependence (for
$p\approx2.2$).

\begin{figure}[!htb]
\centering
\includegraphics[height=8.75cm,width=6.3cm,angle=-90]{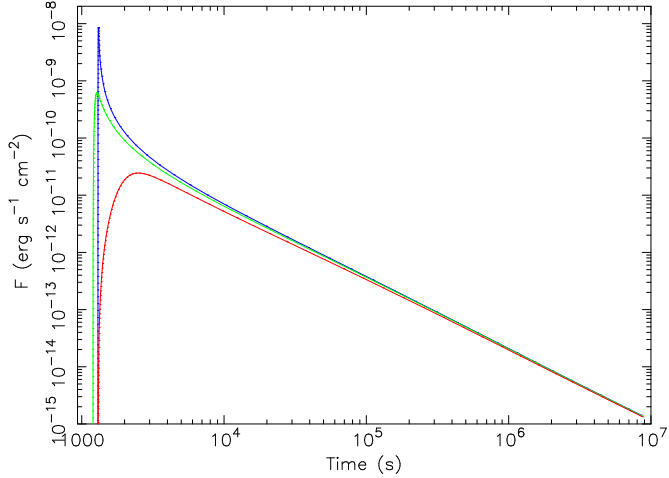}
\caption{ Effects of the initial Lorentz factor $\Gamma_0$ on the X--ray light
curve with the origin of the time shifted to $t_0$=1300 s. The blue curve was
obtained for $\Gamma_0=200$, the green curve for $\Gamma_0=85$, and the red curve for
$\Gamma_0=30$. Note, as the peak in the curve increases with $\Gamma_0$, the other
model parameters are $E_{53}$=0.03, $n$=1, $\varepsilon_e$=0.01, $\varepsilon_B$=0.05,
and $p$=2.2.}
\label{lorentzfactor}
\end{figure}

Figure \ref{lorentzfactor} shows the effects of the initial Lorentz
factor $\Gamma_0$. Its value influences both the height and the
wideness of the peak in the X--ray light curve. The greater 
$\Gamma_0$ is, the higher and narrower the peak is. For $t \gg t_0$,
i.e., when the deceleration phase has been reached for the
different values of $\Gamma_0$, the light curve is independent of
$\Gamma_0$.

\section{External shock from long duration engine activity in XRF 011030 and GRB 011121}
\label{modelapplication}

\begin{figure}[!htb]
\centering
\includegraphics[height=8.75cm,width=6.3cm,angle=-90]{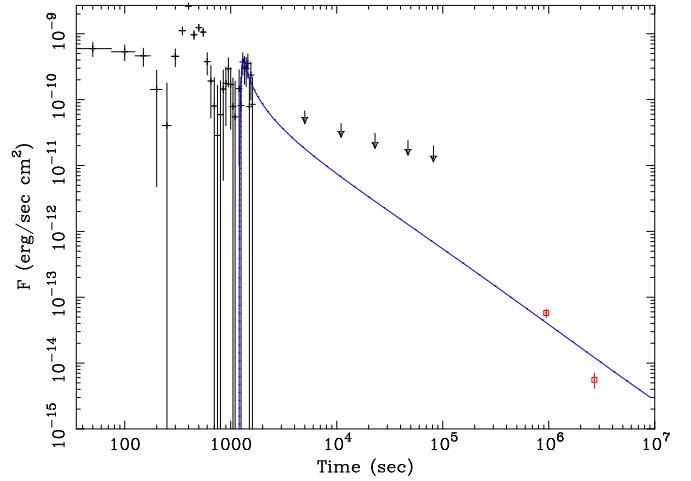}
\caption{X--ray light curve of \xrf in an ISM obtained shifting the origin of 
the time to the onset of the flare, $t_0=1300$s. The model parameters are
$E_{53}=0.03$, $\Gamma_0=110$, $n=1$, $\varepsilon_e=0.2$, $\varepsilon_B=0.05$,
and $p=2.1$.}
\label{ism}
\end{figure}

\begin{figure}[!htb]
\centering
\includegraphics[height=8.75cm,width=6.3cm,angle=-90]{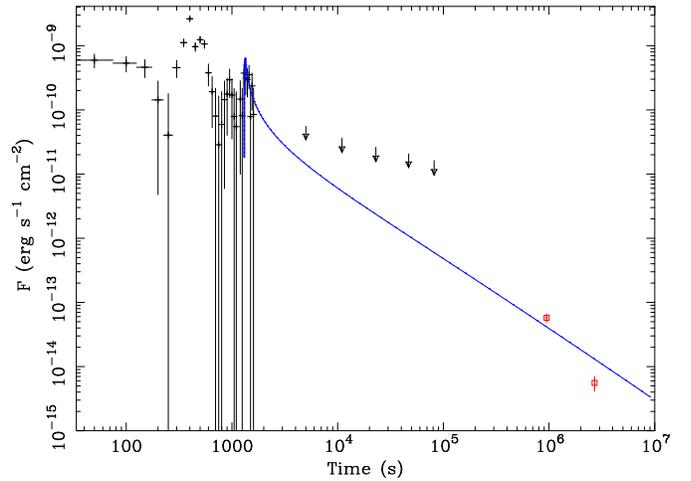}
\caption{X--ray light curve of \xrf in a wind obtained shifting the
origin of the time to the onset of the flare, $t_0=1300 s$. The model parameters are
$E_{53}=0.03$, $\Gamma_0=50$, $A_*=0.05$, $\varepsilon_e=0.1$, $\varepsilon_B=0.05$,
and $p=2.1$.}
\label{wind}
\end{figure}

We applied the model described in the previous section to \xrf and also 
to \object{GRB 011121}. In the case of \object{XRF 011030}, we studied the 
event both in an ISM and in a wind profile environment, producing the 
calculated light curves and finding a family of solutions corresponding to 
several choices of the model parameters. In Figs. \ref{ism} and  \ref{wind} 
we report two possible solutions for a fireball expanding in an ISM and in 
a wind profile environment, respectively, with the origin of the time
shifted to the onset of the flare, $t_0=1300$ s. Small changes
($\lesssim 10$\%) of $t_0$ do not appreciably modify the results.

We find that the  calculated light curves can describe the flare,
both for a fireball expanding in a wind profile environment and
for a fireball interacting with a uniform medium. These two light
curves do not fit the late afterglow data; this will be discussed
in the next section.

\begin{figure}[!htb]
\centering
\includegraphics[width=6.3cm,height=8.76cm,angle=-90]{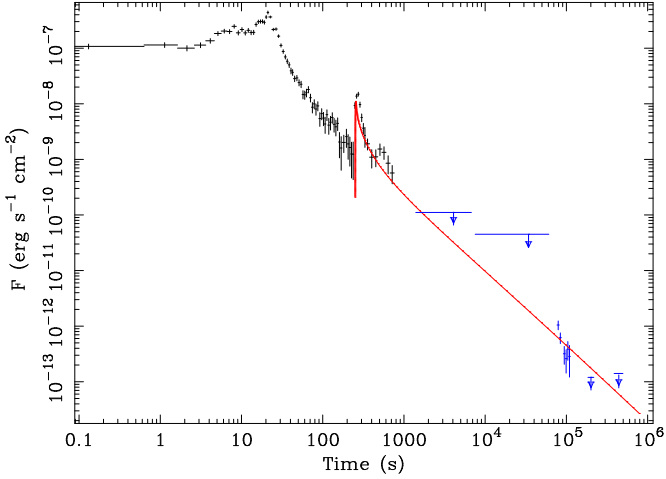}
\caption{ X--ray light curve of \grb for a fireball expanding in a wind with
the origin of the time shifted to the instant of the flare, $t_0=250$ s. The
model parameters are $E_{53}=0.28$, $\Gamma_0=130$, $A_*=0.003$, $\varepsilon_e=0.01$,
$\varepsilon_B=0.5$, and $p=2.5$.}
\label{011121}
\end{figure}

In the case of the flare of \object{GRB 011121}, we computed the light 
curve only for a fireball interacting with a wind density profile. In fact,
\citet{piro05} established that the \grb X--ray and optical data
are consistent with a fireball expanding in a wind environment 
due to the temporal decay observed in these two bands.
Using their parameters and shifting $t_0$ to the onset of the
flare, we find the light curve of Fig. \ref{011121}. The model
describes the flare and the late afterglow, in agreement with the
analysis made by \citet{piro05}.

\subsection{The interpretation of the break in the light curve of XRF 011030}
\label{break}

In Fig. \ref{afterglow}, we note that the backward extrapolation
of the late afterglow flux detected by Chandra is not compatible
with the upper limits observed by BeppoSAX, suggesting the
presence of a temporal break.

First we considered the possibility that the temporal break is
related to a spectral break, i.e., to the passage of the cooling
frequency $\nu_c$ in the X-ray band. We first studied the case of
a wind density profile, when $\nu_c$ increases with the time as
$t^{1/2}$.
Initially, $\nu_c$ can be smaller than $\nu_{obs}$, but there will be 
an instant at which it  becomes greater than $\nu_{obs}$. This marks a 
break in the light curve, which becomes steeper by $\delta \alpha=0.25$.

The observational data of \xrf suggest that the break occurs
between $10^{4}$ and $10^{6}$ s after the burst. We thus require
that $\nu_c$ passes in the X--ray band during this temporal range.
Eq. (\ref{eq:coolwind}) for a given time of the break $T_b$
links  $A_*$ with $\varepsilon_B$.
We derive $\varepsilon_e$ and $\Gamma_0$ values to
reproduce the light curve of the flare as described in the
previous section. $A_*$ is constrained from late X-ray data and
also optical and radio data (see next section). Finally, the
corresponding value of $\varepsilon_B$ is constrained from
Eq. (\ref{eq:coolwind}).
We first attempt to find a broadband solution with the
isotropic energy fixed to $E_{53}=0.03$ (see Sect. \ref{reburst011030}).
In this case, we have problems fitting the radio data. We find a set 
of model parameters able to describe the emission observed in the 
X--ray and optical bands, but the corresponding radio light curve 
is always below the observational data. When the fireball expands 
in a  stellar wind, the flux in the radio band goes as:

\begin{equation}\label{eq:radiofluxwind}
F_{\nu}\propto E_{53}^{1/3}A_*\varepsilon_{e,-1}^{-2/3}\varepsilon_{B,-3}^{1/3}. 
\end{equation}

\begin{figure}[!htb]
\centering
\includegraphics[height=8.75cm,width=6.3cm,angle=-90]{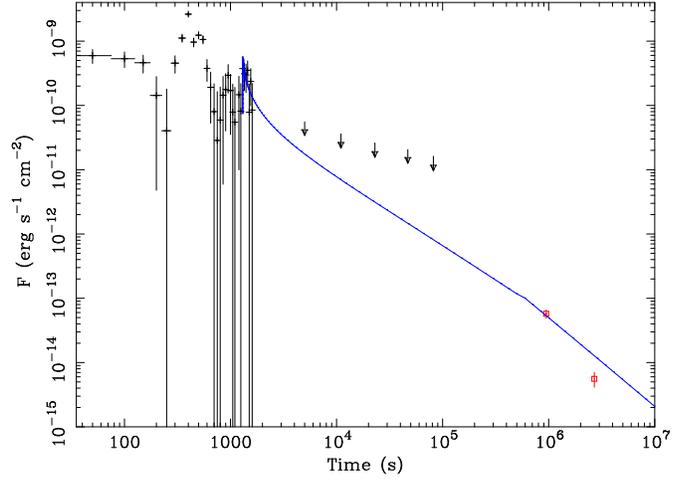}
\caption{ X--ray light curve of \xrf in a wind with the origin of the time
shifted to $t_0=1300 s$. This solution also takes into account the radio and
optical data. The model parameters are $E_{53}=0.3$, $\Gamma_0=60$,
$A_*=0.055$, $\varepsilon_e=0.02$, $\varepsilon_B=0.001$, and $p=2.1$. We have
assumed the efficiency in the conversion of the kinetic energy to be $\eta$=0.1.}
\label{flat}
\end{figure}

The parameters $\varepsilon_e$ and $\varepsilon_B$ are determined 
by X-ray and optical data (see the next Sect. for more detail). Then, 
if we keep $E_{53}=0.03$ to obtain the right normalization of 
the radio light curve, we need to increase the wind density.
On the other hand, this will also cause the normalization of the X--ray 
and optical light curves to increase and surpass the observational 
data. This has motivated us to assume an efficiency $\eta=0.1$ to convert 
the kinetic energy released by the central engine in $\gamma$--rays.
This choice is supported by several authors. \citet{guetta01} and \citet{koba01}
have argued that internal shocks convert the energy with an efficiency 
$\eta \sim 0.1-0.5$, and this was also recently supported by Swift 
observations \citep{zhang05,granot06}. Under this assumption (i.e., 
$E_{53}=0.3$) we find that it is possible to explain the radio data 
jointly with the X-ray and optical data, obtaining a broadband solution.
In Fig. \ref{flat}, we show the calculated X--ray light curve.

It is interesting to note that before the break ($\nu_c<\nu_{obs}$), i.e., during 
the flare, the expected photon index is $\Gamma=(p/2+1)=2.05$, consistent with the 
value found in our spectral analysis (Sect. \ref{analysis}). After the break
($\nu_c>\nu_{obs}$), the expected photon index is $\Gamma=[(p-1)/2+1]=1.55$, 
which agrees with the value $\Gamma=1.72\pm0.20$ found by \citet{vale} in the 
analysis of the Chandra data.

\begin{flushleft}
With regard to the fireball temporal evolution after the break,
$\nu_c>\nu_{obs}$ \citep{panaitescu02}:

\begin{equation}\label{eq:aftertimewind}
F\propto t^{-(3p-1)/4},
\end{equation}
\end{flushleft}
and we expect that the temporal decay index is $\alpha_2=1.325$.
The analysis of Chandra data shows that after the break
$\alpha_2=2.25\pm0.60$ \citep{vale}, these values
are marginally consistent.

Similar considerations can be made for a fireball expanding in an
ISM. In this case, the cooling frequency $\nu_c$ decreases with the
time as $t^{-1/2}$. Supposing $\nu_c>\nu_{obs}$ before the flare,
there will be an instant when $\nu_c$ becomes smaller than
$\nu_{obs}$ and a break occurs.
\begin{flushleft} Now, after the break, the temporal decay is 
slower than the case of a wind density profile \citep{panaitescu02}:

\begin{equation}\label{eq:aftertimeism}
F\propto t^{-(3p-2)/4}
\end{equation}
\end{flushleft}
and the temporal decay index expected after the break is
$\alpha_2=1.075$, not consistent with the Chandra data. Moreover
the spectrum after the break steepens, in disagreement with the
spectral data.

In the ISM case, it is therefore even more difficult than in the wind 
case to explain the late afterglow emission without introducing 
a jet structure. The emission coming from a relativistic shell with jet 
symmetry is similar to the one of a spherical fireball, as long as the 
observer is on the jet axis, and the jet Lorentz factor $\gamma$ is greater 
than the inverse of its angular spread $\theta_0$ \citep{rhoads97}. During
its expansion, the fireball collects a growing amount of matter;
thus, the Lorentz factor $\gamma$ decreases and there is an instant
when $\gamma<\theta_0^{-1}$. At this time, the sideways spread of
the jet becomes important and the observed area grows more quickly. 
This leads the flux to decrease more rapidly whit respect to the
spherical case, and we expect a break in the light curve
\citep{sari99}. \citet{sari99} calculated that at high frequencies
the flux decreases like $t^{-p}$ both when $\nu_{obs}>\nu_c$ and
when $\nu_{obs}<\nu_c$.
Thus, with the electron population index $p=2.1$, the predicted
temporal behaviour agrees with the two Chandra observations.
Once the sideways expansion of the jet becomes important, the
cooling frequency $\nu_c$ is constant with time \citep{panaitescu01} 
and the spectrum should not evolve. In our data, the \xrf spectral 
evolution is only marginally significant; in fact, the photon index 
$\Gamma$ of the power law fitting the flare is consistent within the 
errors with the photon index of the power law describing the afterglow 
(Table \ref{analysis}).
We therefore carried out a comparison between the model and broadband
data, i.e., taking into account the optical and the radio information
discussed in the next section.

\begin{figure}[!htb]
\centering
\includegraphics[height=8.75cm,width=6.3cm,angle=-90]{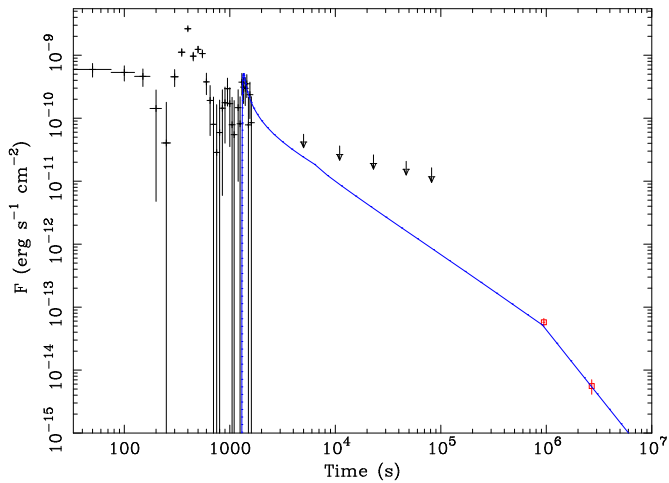}
\caption{X--ray light curve for a jetted fireball expanding in an ISM with
the origin of the time shifted to $t_0=1300 s$. The model parameters
are $E_{53}=0.03$, $\Gamma_0=130$, $n=5$, $\varepsilon_e=0.29$,
$\varepsilon_B=8\cdot10^{-5}$, $p=2.1$, and $T_b=8 \cdot 10^5 s$.}
\label{ismjet}
\end{figure}

In this case, we find a solution that nicely describes all the
data without requiring an efficiency $\eta$ in the conversion of the
kinetic energy in $\gamma$-rays (see  Fig. \ref{ismjet} for the X--ray
light curve).
We notice that even if the jet model has an additional free parameter
with respect to the spherical fireball model for a jetted fireball,
the model parameters are, still well-constrained.
This is mostly due to the passage of the cooling frequency $\nu_c$ in
the X--ray band.
At the start of the observation $\nu_c>\nu_{obs}$, at about $10^4$
s, the cooling frequency $\nu_c$ becomes smaller than $\nu_{obs}$.
After this instant the X--ray and optical flux follow the same law, and this
well constrains the model parameters.
The constraints on the model parameters given by optical and radio information
are discussed with more detail in the next section.

\section{Broadband analysis of XRF 011030 afterglow data}
\label{optical}

No optical counterpart has been detected for \object{XRF 011030}.
From among all the optical observations, we considered those performed
by \citet{vijay} and \citet{rhoads01} because they are the most
constraining.
The upper limits are R$>$21 \citep{vijay} and R$>$23.6 \citep{rhoads01}
at 0.3 and 2.7 days after the burst, respectively. We correct
magnitudes for the reddening due to the absorption of our Galaxy,
finding $R>$ 20.4 and $R>$ 23.1 (corresponding to an optical flux
$F_{\nu, opt1}=1.79\times10^{-28}$ erg cm$^{-2}$ s$^{-1}$
Hz$^{-1}$ and $F_{\nu, opt2}=1.69\times10^{-29}$ erg cm$^{-2}$
s$^{-1}$  Hz$^{-1}$) for the two observations. In the radio band,
\citet{taylor} associated a transient source with a flux 
$F_{\nu,R}=1.81\times10^{-27}$ erg cm$^{-2}$ s$^{-1}$ Hz$^{-1}$ 
about 10.5 days after the burst with \object{XRF 011030}.

\begin{figure}[!htb]
\centering
\includegraphics[height=8.75cm,width=6.3cm,angle=-90]{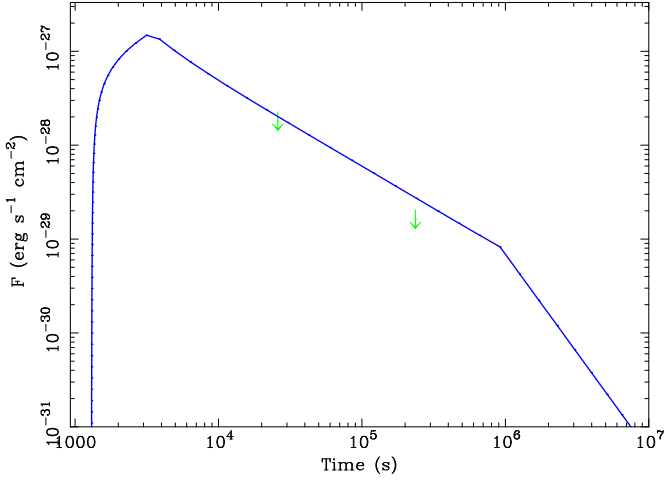}
\caption{ Optical light curve of a jetted fireball expanding in an ISM
with the origin of the time shifted to $t_0=1300$ s. The model
parameters are $E_{53}=0.03$, $\Gamma_0=130$, $n=5$, $\varepsilon_e=0.29$,
$\varepsilon_B=8\cdot10^{-5}$, $p=2.1$, and $T_b=8 \cdot 10^5$ s.}
\label{ismjetottico}
\end{figure}

\begin{figure}[!htb]
\centering
\includegraphics[height=8.75cm,width=6.3cm,angle=-90]{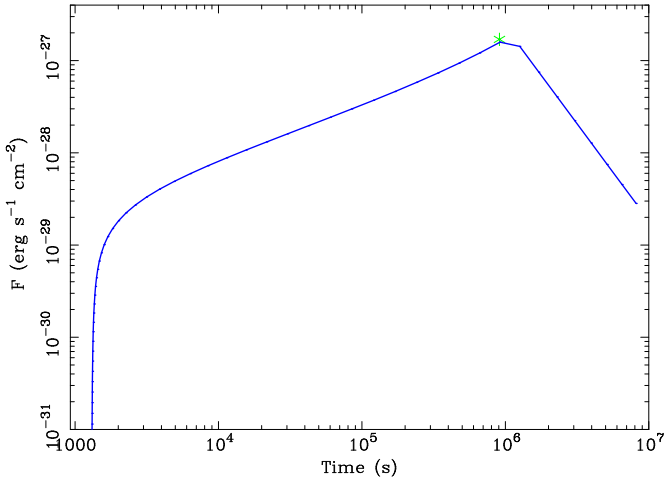}
\caption{ Radio light curve of a jetted fireball expanding in an ISM with
the origin of the time shifted to $t_0=1300$ s. The model parameters
are $E_{53}=0.03$, $\Gamma_0=130$, $n=5$, $\varepsilon_e=0.29$,
$\varepsilon_B=8\cdot10^{-5}$, $p=2.1$, and $T_b=8 \cdot 10^5$ s.}
\label{ismjetradio}
\end{figure}

For a jetted fireball expanding in an ISM, a solution that
accounts for X--ray (Fig. \ref{ismjet}), optical (Fig. \ref{ismjetottico}),
and radio (Fig. \ref{ismjetradio}) is given by $E_{53}=0.03$, $\Gamma_0=130$,
$n=5$, $\varepsilon_e=0.29$, $\varepsilon_B=8\cdot10^{-5}$, $p=2.1$, and
$T_b=8 \cdot 10^5$ s. We show the optical and radio light curves
corresponding to this set of model parameters in Figs, \ref{ismjetottico}
and \ref{ismjetradio}, respectively.
We investigated how well the parameters are constrained, with
particular regard to the density. 
The density is mostly constrained by the data below the cooling frequency,
in this case optical and radio, and whether times are greater than about
$10^4$ s (that is the time when a spectral break occurs), also X--ray data.
After $10^4$ s, the emitted flux in the X--ray and optical
band is given by \citep{panaitescu00}:

\begin{equation}\label{eq:opticalflux}
F_{\nu}\propto E_{53}^{(p+3)/4}n^{1/2}\varepsilon_{e,-1}^{p-1}\varepsilon_{B,-4}^{(p+1)/4}. 
\end{equation}

This same relation applies for the radio data because, at the
time of the observation, the injection frequency $\nu_i$ is below
or very near the observational frequency $\nu_{obs}$.
Thus, the normalization of the light curve in one of the
three observational bands also determines the normalization of the light
curve in the other two bands. This causes the model parameters 
to be well-constrained.

\begin{figure}[!htb]
\centering
\includegraphics[height=8.75cm,width=6.3cm,angle=-90]{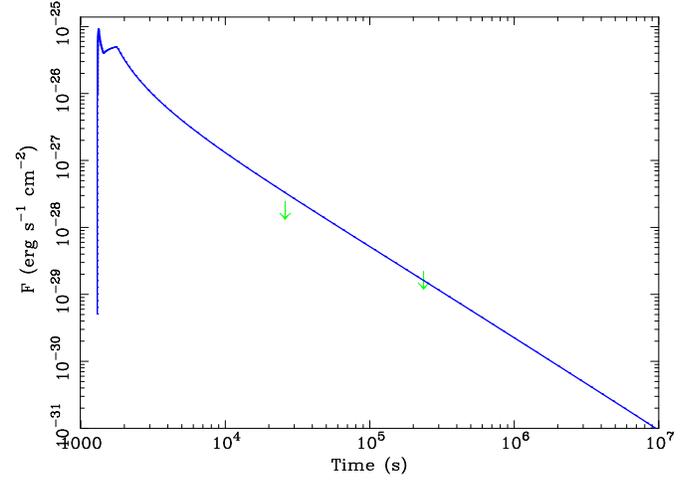}
\caption{ Optical light curve of a spherical fireball expanding in a wind
with the origin of the time shifted to $t_0=1300$s. The model parameters 
are $E_{53}=0.3$, $\Gamma_0=60$, $A_*=0.055$, $\varepsilon_e=0.02$,
$\varepsilon_B=0.001$, and $p=2.1$. We have assumed the efficiency in the 
conversion of the kinetic energy to be $\eta$=0.1.}
\label{windottico}
\end{figure}

\begin{figure}[!htb]
\centering
\includegraphics[height=8.75cm,width=6.3cm,angle=-90]{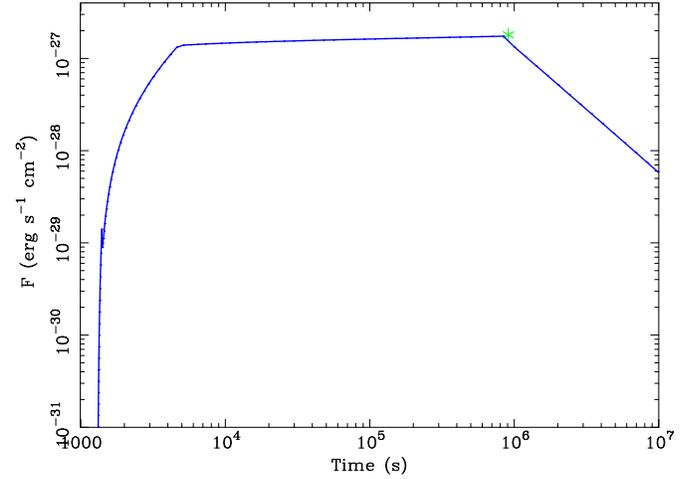}
\caption{ Radio light curve of a spherical fireball expanding in a wind
with the origin of the time shifted to $t_0=1300$s. The model parameters
are $E_{53}=0.3$, $\Gamma_0=60$, $A_*=0.055$, $\varepsilon_e=0.02$,
$\varepsilon_B=0.001$, and $p=2.1$. We have assumed the efficiency in the 
conversion of the kinetic energy to be $\eta$=0.1.}
\label{windradio}
\end{figure}

In the case of the spherical fireball expanding in a wind, the
break has to be self-consistently described (i.e., without the 
addition of a free parameter).
Consequently, the model parameters are also well-constrained, 
$E_{53}=0.3$, $\Gamma_0=60$, $A_*=0.055$, $\varepsilon_e=0.02$, 
$\varepsilon_B=0.001$, and $p=2.1$.
The corresponding X--ray, optical, and radio light curves are shown 
in Figs. \ref{flat}, \ref{windottico}, and \ref{windradio}.

\section{Summary and conclusions}
\label{conclusions}

In this paper, we have presented the temporal and spectral analysis 
of the X-ray flash \xrf observed by BeppoSAX. This event is 
one of the longest in the BeppoSAX sample \citep{zand}, 
with a duration of $\sim 1500$ s. In particular, along with the main 
pulse, we find a precursor event and a late X-ray flare.

While the spectrum of the main burst is not consistent with a
black body, we cannot exclude this model for the precursor. This
result could be due to the lower statistics available, but it is
nonetheless interesting to note that, so far, there was only one
example of a precursor consistent with such a model \citep{murakami}.
This feature could be associated with the cocoon formed by the jet
emerging at the surface of the collapsing massive star
\citep{ramirez,waxman}.

After the launch of the Swift satellite, X--ray flares appear 
to be a common feature in GRBs light curves. This has favored the 
development of a large number of models to explain X--ray flares' origin,
both in internal and external shock scenarios \citep{zhang05}.
X--ray flares observed by BeppoSAX in \grb and in \object{XRR 011211}
have spectra similar to that of the late afterglow.
This similarity can be straightforwardly accounted for in the framework
of the external shock, i.e., the flare represents the onset of
the afterglow and it is connected with the late afterglow emission with
a power law. For \grb and \object{XRR 011211}, a connection with the
late afterglow is acceptable only by shifting the origin of the
time $t_0$ to the instant of the flare \citep{piro05}.
This implies a long duration engine activity (thick shell fireball,
\citet{sari}).

We find that this scenario fits nicely with the observational
data, including the X-ray flare and the late broadband radio,
optical, and X-ray afterglow observations of \xrf. The latter,
performed by Chandra, indicate the presence of a temporal break
occurring between $10^4$ s and $10^6$ s after the burst.
We carried out a detailed modelling of the data, finding good
agreement with observations for a spherical fireball expanding in
a wind medium and for a jetted fireball expanding in an ISM. In the
first case, the temporal break is explained by the passage of the
cooling frequency in the X-ray band.
We cannot exclude that the flare observed in \xrf is
due to internal shocks. However, this would require an engine
activity that is tuned to track the peak energy of the emission
during the prompt and flaring phases.

In the context of the external shock scenario, the time shift
$t_{0}\sim 1300$ s is due to a long lasting central
engine activity that remains active until the time of the flare,
with the most of the energy released at the end of the
emission phase \citep{sari,lazzati}.
In this case, the peak of the afterglow emission coincides with
the flare, and the afterglow decay will be described by a self-similar
solution by counting the time from the instant the inner engine turns
off, i.e., when $t_0\simeq t_{eng}$.

We also find that a thin shell fireball cannot describe the flare
in a continuous density profile, in fact, in this case, the flux
rises and decays too slowly to describe the shape of the flare. Also,
with a discontinuous density profile, it is very difficult to produce 
flares as high and narrow as that observed in \object{XRF 011030}, 
unless one assumes that the fireball is expanding in a clumpy 
medium \citep{dermer99,dermer05}.

Recently, Swift observations have shown the presence of late X-ray
flares in several other events. Some of these flares appear to
have a spectral behaviour consistent with that of late X--ray
flares observed by BeppoSAX, i.e., a soft spectrum substantially
well-differentiated from the hard, prompt emission typically
attributed to internal shocks (\object{GRB 050126}, \object{GRB
050219a} \citep{tagliaferri05}, and \object{GRB 050904}
\citep{burrows_rew}).
\object{GRB 050126} and \object{GRB 050219a} appear to
follow a $(t-t_0)^{-\alpha}$ power law, with $t_0\approx 100$ s, 
reasonably well. Instead in other flares, e.g \object {XRF 050406}
and \object{GRB 050502B}, the hardness ratio suggests a spectral
evolution resembling the prompt emission \citep{burrows_rew}. This
behaviour has been thus interpreted by \citet{burrows} as due to
internal shocks produced by a long duration central engine activity.
However we notice that in some cases it could be possible
explain hard-to-soft spectral evolution also in the context of the
external shock scenario.
When the fireball expands in a wind the cooling frequency $\nu_c$
increases with the time and there will be an instant when it enters
in the X--ray band. At this instant the spectral index changes from
$\beta=p/2$ to $\beta=(p-1)/2$ and the spectrum becomes harder of a
factor 0.5.

Swift observations showed also the presence of multiple flares 
in GRB light curves on relatively small time scales, from $\sim$ 100 s 
up to $\sim$ 1000 s. For example \object{GRB 050421} shows two successive
flares within about 150 s \citep{godet}, \object{GRB 050607} has two
flares within 500 s \citep{pagani} and \object{GRB 050730} has three
flares within 800 s \citep{burrows_rew}.
In these cases a thick shell fireball can be successful to explain only
one of the flares appearing in the light curve, i.e. only one flare can
be identified with the beginning of the afterglow emission.
The other flares can be attributed to internal shocks or to the
interaction with a clumpy medium in the framework of the external shock
scenario.
In conclusion we note that the present data suggest the existence
of two categories of late X--ray flares differentiated by their
spectral behaviour.
It is also interesting to note that both in the framework of the
internal shocks scenario and in that of external shocks, the
explanation of late X--ray flares requires a central engine that
remains active about until the time of the flare.

\begin{acknowledgements}

The authors are grateful to E. Massaro, V. D'Alessio, B.Gendre and
A. Corsi for useful discussions and comments.  This work was
partially supported by the EU FP5 RTN Gamma ray bursts: an enigma
and a tool. The BeppoSAX satellite was a  program of the Italian
space agency (ASI) with participation of the Dutch (NIVR) space
agency.

\end{acknowledgements}

\end{document}